# Correlative extinction and single fluorophore bleaching microscopy for ligand quantification on gold nanoparticles

Nicole Slesiona,*[a] Lukas Payne,[a,b] Iestyn Pope,[a] Paola Borri,[a] Wolfgang Langbein,[b] Peter Watson[a]

[a] School of Biosciences, Cardiff University, Museum Avenue, Cardiff CF10 3AX, United Kingdom
[b] School of Physics and Astronomy, Cardiff University, The Parade, Cardiff CF24 3AA, United Kingdom

**Abstract** - Nanoparticles (NPs) are promising therapeutic delivery agents, yet it is increasingly apparent that the number and manner of presentation of cell binding ligands on the NP can affect the eventual fate of the therapeutic. Whenever NPs are conjugated with biomolecules, a heterogenous population of decorated NPs will be produced and the details of the subpopulations of particle-ligand structures needs to be characterised for a reliable interpretation of NP-based data. We report an optical microscopy method to quantitatively evaluate the conjugation on a single particle basis in samples consisting of gold NPs (GNPs) decorated with human holo-transferrin fluorescently labelled with Alexa647 (Tf). We employed widefield fluorescence and extinction microscopy on NP-ligand constructs sparsely deposited onto a glass surface, alongside a correlative analysis which spatially co-localises diffraction-limited sources of fluorescence with the optical extinction by individual GNPs. A photobleaching step analysis of the fluorescence emission was employed to estimate the number of fluorophores contributing to the detected emission rate. The method quantifies the number of fluorescent biomolecules attached per GNP, the numbers of unconjugated GNPs and unbound Tf present within the mixed population, and the size and intraparticle clustering propensity of conjugated GNPs. We found a high variability in the number of Tf ligands per GNP within the GNP population, when analysed at the single-particle level, unraveling a non-trivial statistical distribution not accessible in ensemble averaged approaches.

## 1. Introduction

Metal nanoparticles (MNPs) have generated ever-increasing attention as they have become an integral part in catalysis[1–3], biomedical detection[4–6], and nanotherapy[7,8,9] Special attention is laid on gold nanoparticles (GNPs) due to their low cytotoxicity, high biocompatibility, chemical stability, and their unique optical properties in the visible range of light that depend on their size and shape[10,11]. Their surface chemistry and geometrical characteristics can either enhance or hamper their specificity and reliability for the intended application[12,13]. Both properties can be controlled as GNPs can be synthesised in a variety of sizes and shapes with a narrow size distribution in a single-phase reaction. Biofunctionalization of GNPs is a crucial step in the creation of nanocarriers with targeted sensory and/or therapeutic relevance and requires careful consideration of the involved reagents to ensure colloidal stability at a range of pH levels and salt concentrations of their surrounding medium. Generally, GNP are coated with bifunctional polymers such as polyethylene glycol (PEG) to stabilize them against aggregation and to introduce chemical moieties for their subsequent biofunctionalization. The surface density of the stabilising capping agent will govern their stability and half-life in the body, while the nature and density of targeting molecules (TMs) controls specific binding. The density of the latter may control the uptake mechanisms induced by the nanocarrier, as this property - in conjunction with the GNP shape, level of curvature, and size – influences the way TMs are presented to an analyte or a biological membrane.

Both GNP size and TM surface density need thorough evaluation to allow the correct interpretation of GNP-based data, their use as imaging agents, and transition into clinical applications. Generally, current quantification methods on MNPs begin with the determination of the total particle surface area by gas sorption analysis[14,15] or by measuring the average MNP size of the ensemble by dynamic light scattering (DLS)[16,17], transmission electron microscopy (TEM)[18–20], or scanning electron microscopy (SEM)[19]. Subsequently, the total number of TMs in the sample is determined. Quantitative methods include thermogravimetric analysis (TGA)[20,21], optical spectroscopy[22], detaching the TMs from the particle and measuring their concentration in solution[23,24], by nuclear magnetic resonance (NMR)[20,25] or vibrational spectroscopy[26]. Emerging techniques such as pH-based methods, electrospray-differential mobility analysis, and X-ray photoelectron spectroscopy (XPS) also show promising results for the field of TM quantification[20,27]. The average particle size and the total number of TMs in the sample are then used to calculate the average TM surface density on an average sized MNP in the ensemble. Knowing this average density may be sufficient to judge whether a conjugation reaction has been successful, however further information on the heterogeneity of the population is needed to correctly interpret experiments. Functionalization reactions of NPs with TMs inevitably produce a polydisperse sample. Apart from unconjugated NPs and unconjugated TMs, the sample can contain NPs with varying biomolecule number and spatial distribution, aggregates of NPs crosslinked by the TMs, as well as TM aggregates. Indeed, there have been studies revealing missing colocalization between NPs and fluorescent TMs thought to be attached on the NP surface.[28,29] This polydispersity makes interpretation of cell uptake with TM functionalised NPs challenging. For example, pure TM aggregates will generate a fluorescence signal that could be interpretated as a NP bound to the cell membrane, whilst cross-linked NPs may reach a size too large to be internalized by mammalian cells[12,30].

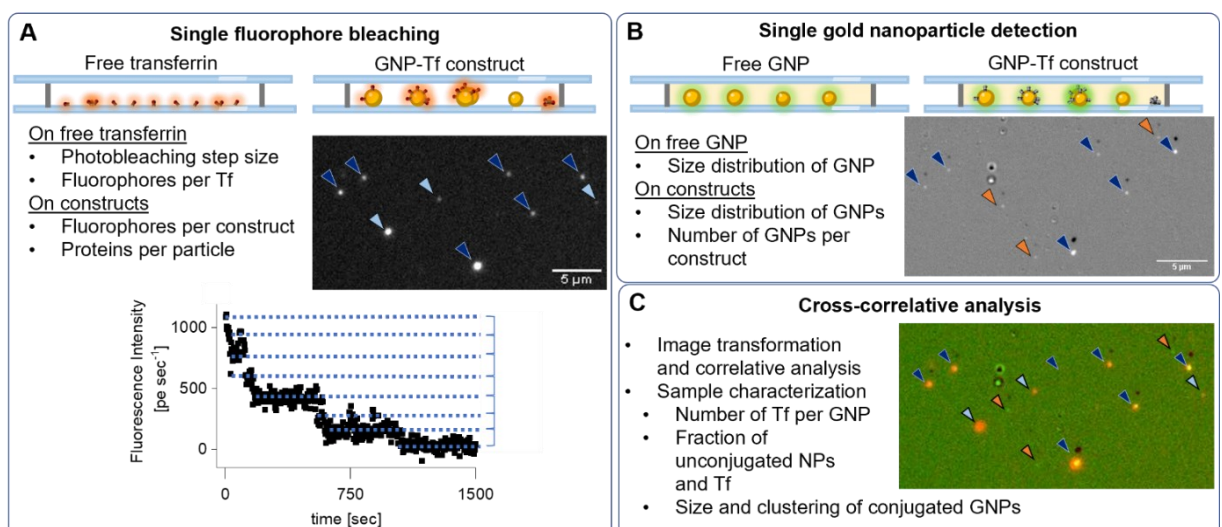

**Figure 1 Methodology of correlative measurements of single fluorophore bleaching and extinction microscopy.** NP constructs consist of spherical GNPs decorated with fluorescently labelled human *holo*-transferrin (Tf). A The fluorescence intensity of individual free Tf, as well as of GNP-Tf constructs, spin-coated onto a glass surface is imaged by widefield microscopy over time until all fluorophores are bleached. The bleaching step size per fluorophore and in turn the number of fluorophores per protein is determined. The fluorescence intensity of GNP-Tf constructs is then used to quantify the number of Tf contributing to an individual GNP. Fluorescence spots (not) co-occurring with extinction spots are marked with dark blue (light blue) arrows. B The extinction cross-sections of particles in the area that was previously photobleached are quantified using widefield extinction microscopy. The number of GNPs contributing to an individual spot can be estimated from the average extinction cross section of individual particles. Extinction spots (not) co-localised with a fluorescence spot are marked with a dark blue (orange) arrow. C Fluorescence and extinction images are overlaid using a transformation that accounts for shift, scaling, rotation, and shear between images acquired in different microscopes, to allow evaluation for co-localisation of fluorescence and extinction spots.

At present, there is no method to evaluate the parameters quantifying a sample's polydispersity of NP size and number of TMs binding to the NP on a particle-by-particle basis. For nanoparticles we are limited to ensemble quantification methods that evaluate the number of biomolecules per particle on average[31].

In this work, we present a method that allows quantification of TM numbers on GNPs on a particle-by-particle basis. This is realised by a correlative analysis of fluorescence and extinction microscopy images of GNP constructs, here using functionalized spherical GNPs decorated with the fluorescently labelled ligand human *holo*-transferrin (Tf). Single molecule bleaching data reveal the number of fluorophores contributing to the fluorescence of a single Tf, and in turn quantify the number of Tfs per GNP (Figure 1A). Wide-field extinction microscopy allows the determination of the size and the estimated number of GNPs contributing to the extinction spots (Figure 1 B). Extinction microscopy is a method to quantitatively measure extinction cross-sections of hundreds of MNPs simultaneously using commonly available optical components. Absolute extinction cross-section values are then retrieved by an automated analysis, which allows a high-throughput characterisation of NP sizes[32,33]. An image transformation software is used to create matching fluorescence and extinction images of the same area to analyse the co-occurrence of extinction and fluorescence signals. The analysis produces a quantifiable output in respect to the number of attached Tf per GNP, the fraction of successfully functionalised GNPs *vs.* unconjugated GNPs *vs.* free Tf in a sample, and the number of GNPs that have been cross-linked by the reaction procedure (Figure 1 C). The method is also capable of generating a measure for how many Tfs have been crosslinked without a GNP present in the aggregate. These aggregates especially might be misinterpreted as successfully internalised particle-ligand constructs in fluorescence-based microscopy measurements. Notably, our analysis can provide information to guide the optimisation of the functionalisation reaction using for example different concentrations of ligands, cross-linkers, and particles.

## 2. Methods

### 2.1 Samples

#### 2.1.1 GNP functionalization with Tf

Three GNP-Tf constructs were utilized in this work, differing in their ratios of proteins to attachment sites on the GNP (see Table 1). Commercially available GNPs

(Nanopartz™ Ntracker™, Salt Lake City, UT) were used with 14 nm (sample 1) and 20 nm (sample 2 and 3) average diameter and a proprietary coating of stabilizing hydrophilic polymer with terminal amine groups (2 attachment sites per nm$^2$ surface area, specified by the manufacturer). For conjugation with commercially available Alexa647-labelled human holo-transferrin (TfA647, Thermo Fischer, Loughborough, Leicestershire, UK), sulfo-succinimidyl-4-(N-maleimidomethyl)-cyclohexane-1-carboxylate (sulfo-SMCC) was used. Ratios of protein to available attachment sites on the GNPs were formulated as follows: 0.01 Tf per attachment site in sample 1, 0.4 Tf per attachment site in sample 2, and 1 Tf per attachment site in sample 3. This was realised by diluting the GNP concentration and keeping the concentration of Tf constant at 340 nM. First, the GNPs were conjugated to the NHS functional group of the crosslinker by adding a 100x excess of sulfo-SMCC to amine residues. The reaction was left on a shaker at room temperature for an hour after which the GNPs were purified by centrifugation at 12,000 rcf for 20 min. About 90% of the supernatant were removed and replaced with PBS (pH 7) containing 340 nM Tf. The reaction was left for 2h at room temperature on a shaker while protecting the samples from light. After incubation, to remove unbound Tf the sample was purified by centrifugation at 12,000 rcf for 20 min, 90% of the supernatant removed and discarded, and the particles resuspended to a volume of 1 mL 1% (v/v) PBS 0.01% (v/v) Tween 20 solution. The purification step was applied three times, and after the last step of supernatant removal, GNPs were resuspended to a final volume of 1 mL with PBS. The functionalized GNPs were stored at 4°C until usage for a maximum of one month.

### 2.1.2 Sample preparation for imaging

A 24x24 mm (thickness 0.13-0.17 mm, Menzel-Gläser, Braunschweig, Germany) H$_2$O$_2$-cleaned cover slip was placed on the chuck of a spin coater (WS-650MZ-23NPPB, Laurell, USA) and immobilized by vacuum. The disk was set to spin 35 sec at 2000 rpm followed by another 30 sec at 4000 rpm. During the first 5 sec of rotation, 20 μL of sample were pipetted onto the center of the spinning cover slip. Given that single Tfs and GNPs are smaller than the resolution achievable on diffraction-limited optical imaging systems, fluorescence signals from individual Tfs or GNP-Tf constructs appear as spots with a size given by the point spread function (PSF) of the microscope. The size of the PSF is a function of the numerical aperture (NA) of the objective used, the illumination wavelength, and the refractive index of the surrounding medium. Where two nano-objects are much closer than the size of the PSF, they will not be separated and appear as a single object. To minimise the likelihood of such events, the Tf, the GNP, and the GNP-Tf constructs were diluted to a density of 10$^9$/mL before spin-coating. Considering the 20 μL of sample used and a cover slip surface area of 576 mm$^2$, this corresponds to 0.035 particles/μm$^2$ (average particle distance 5 μm) when assuming a uniform coating of the cover slip. Note, that before the spin-coating, a diamond scribe was used create recognizable marks on the sample glass surface to allow measurements of the same area on different microscope platforms. The sample was then mounted, onto a glass slide, spin-coated surface towards the slide, using a 0.12 mm thick adhesive gasket (Grace Bio-lab SecureSeal) modified with channels to allow the injection of index matching oil for extinction microscopy.

## 2.2 Data acquisition and processing

### 2.2.1 Wide-field epi-fluorescence microscopy and data processing

Wide-field epi fluorescence measurements were conducted on an inverted Olympus IX73 microscope and a Prior Lumen200Pro light source using filter set (89000, Chroma, Vermont, U.S.A.) selecting the ET645 nm/30 nm as exciter filter and the ET705 nm/72 nm as emitter filter. The emission was detected with a sCMOS Camera (Hamamatsu ORCA-flash 4.0 V2, 30,000 full well capacity, 1.4 electrons read noise, and 0.46 electrons per count) at a 2 × 2 binning readout on all images corresponding to a 13 μm pixel size. Camera and filter settings were operated utilizing the HCImage software package (Hamamatsu). A 100× oil immersion objective with an adjustable NA of 0.6-1.3 was used. The objective NA was set to 0.95 to avoid total internal reflection which was found to create a significant background by residual transmission of the filters. An aperture iris was not available in the fly-eye illuminator of the IX73, which would have enabled limiting the NA of the illumination while collecting the emission with the highest NA for maximum signal. To determine the number of fluorophores per Tf and the number of Tf per GNP, spin-coated samples of Tf and GNP-Tf were measured in a time series with 3 sec integration time. To define the excitation intensity, a red fluorescent slide (ThorLabs - FSK6), was mounted and the lamp intensity adjusted to achieve an average signal count of 50,000 counts at 50 ms exposure time. To minimize photobleaching prior to measurement the sample was focussed in bright field at minimum light intensity. To remove background signal and hot pixels, a background image was acquired using the same acquisition settings with the fluorescence excitation blocked, using an average of 20 frames. These background images were subtracted from the time-series prior to analysis with the analysis software. The data was evaluated using an image analysis software suite developed in-house (Extinction Suite Macro[33,34]) to extract the fluorescence intensity of an individual spot, spatially summed over the PSF, versus time to determine steps of photobleaching. For each fluorescent spot the data was summed over a 4 pixel radius around its peak and a local background from the surrounding data up to 8 pixels radius was subtracted. The suite assigns an index to each spot to allow cross-referencing to its corresponding extinction spot for co-occurrence analysis. Details of the analysis software and its calculations are discussed in Payne et al.[33]. Steps of photobleaching were evaluated manually and the resulting distributions plotted to quantify the average number of fluorophore ligands present per individual Tf, subsequently used to determine the number of Tf per GNP.

Table 1 Gold nanoparticle-transferrin construct samples used in the experiments.

| Sample | 1 | 2 | 3 |
|---|---|---|---|
| mean particle diameter [nm] | 14 | 20 | 20 |
| attachment sites per particle | 1230 | 2500 | 2500 |
| ratio of attachment sites to Tfs | 0.01 | 0.4 | 1 |

### 2.2.2 Wide-field extinction microscopy

Following photo bleaching measurements, the chamber of the sample was filled with silicone oil, index matched to glass (n = 1.52), to reduce background caused by surface roughness in extinction measurements. Extinction microscopy was performed as described by Payne et al.[32]. Briefly, measurements were performed on a Nikon Ti-U inverted microscope, using a 100x 1.45 NA planapochromat (Nikon lambda series, MRD01905) oil objective with a 1x tube lens, and a 1.34NA condenser. The sample was illuminated by a halogen tungsten lamp (V2-A LL 100 W; Nikon) with a bandpass filter in the illumination path of 550 ± 20 nm (Thorlabs FB500-40). Extinction images were obtained with a scientific-CMOS (sCMOS) camera (PCO Edge 5.5), with 2560 × 2160 pixels and 16-bit digitization, 0.54 electrons per count, and a full well capacity of 30 000 electrons. The exposure time was set to 12 ms for each frame, using a 82 Hz frame rate and an average signal of about 22,000 photoelectrons (pe). To acquire a differential transmission image, the area of interest was imaged twice with the particles in focus, with a lateral sample shift of 1.3 µm. Background images were acquired by blocking the illumination path. To achieve the required sensitivity in extinction cross-section, 1280 frames were acquired for each of the shifted positions. To reduce the effect of slow drifts in illumination and sensor, only 128 frames were sequentially acquired at fixed sample position, and the position was cycled 10 times. The resulting $\sigma_{ext}$ data had a noise of about 15 nm². An extinction image was calculated using the averages of all frames at each respective position as described in Payne et al.[32] and was conducted with the same analysis software that was used for fluorescence intensity evaluation. Also, here an index is assigned to each spot detected by the software so they can be cross-referenced to their corresponding fluorescence spot.

### 2.3 Correlative image analysis

All images were evaluated using ImageJ. Because extinction and bleaching measurements of the same areas were acquired on different microscope systems, extinction images were transformed onto the fluorescence images to correct for scaling, shear, rotation and translation differences, using an in-house written image transformation software. Based on a number N of common features identified by their coordinates of fluorescence ($\mathbf{r}_f$) and extinction ($\mathbf{r}_e$) signals, the software determines a transformation

$$\mathbf{r}_f = C\mathbf{r}_e + \mathbf{t}$$

where $\mathbf{t}$ is a translation vector, and the matrix $C$ rotates, shears, and scales the axes. The transformation parameters are determined by minimizing the standard deviation $S$ between $\mathbf{r}_f$ and the transformed $\mathbf{r}_e$, given by

$$S = \sqrt{\frac{1}{N}\sum |\mathbf{r}_f - (C\mathbf{r}_e + \mathbf{t})|^2}.$$

At the minimum of N=3 required, the transformation is exact, so that S=0. The transformed image that can be overlapped with its reference image for correlation analysis (Figure 1 C). Details of the registration method are described in Pope *et al.* 2022[35], Overlapped images were evaluated manually to assess co-localisation of extinction and fluorescence signals.

## 3. Results and discussion

### 3.1 Determination of fluorophore bleaching step size

Free Tf and GNP-Tf construct samples were imaged on a wide-field microscope in epifluorescence (see Methods section 2.1 and 2.2.1) to measure the fluorescence rate over time for the evaluation of the mean bleaching step size $I_B$. Its value allows to calculate the number $N_F$ of fluorophores per Tf (or GNP-Tf construct), by dividing the fluorescence rate $I(t_0)$ detected for an individual spot in the image at the initial time point $t_0$ in the sequence, by $I_B$, so that

$$N_F = \frac{I(t_0)}{I_B}.$$

$I_B$ was determined for fluorophores attached to free Tf (number of analysed Tf (number of analysed spots N = 91) and GNP-Tf constructs (N = 95), see Supplementary Figure S2. Free Tf showed a mean $I_B$ of 150 ± 79 pe per second (pe s$^{-1}$), and GNP-Tf a mean $I_B$ of 142 ± 48 pe s$^{-1}$.

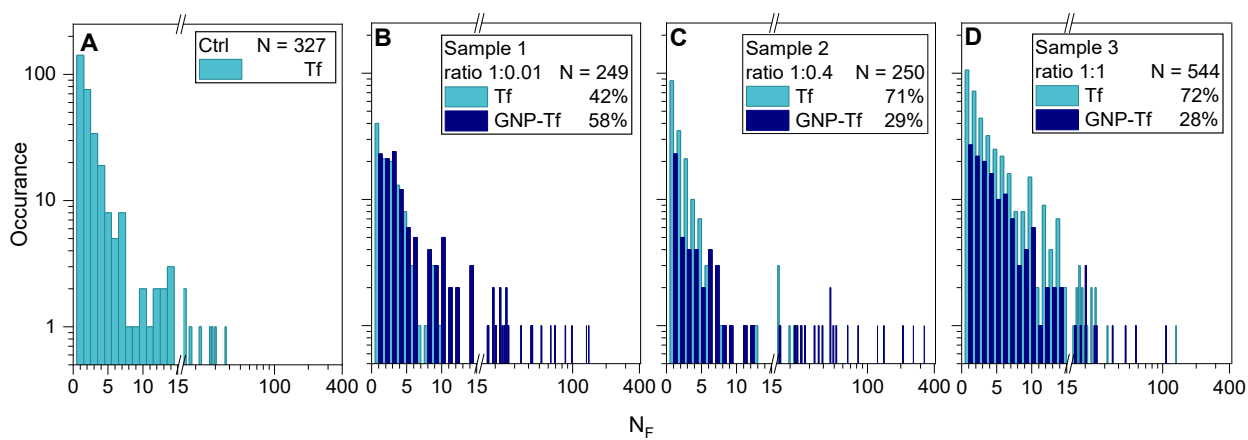

**Figure 2 Distributions of fluorophore counts on free Tf and GNP-Tf constructs.** Distribution of $N_F$ for free Tf (light blue) and GNP-Tf constructs (dark blue) on a sample containing free Tf (A), and on GNP-Tf sample 1 (B), 2 (C), and 3 (D). Fractions of GNP-Tf represent fluorescent particles only i.e. do not include non-fluorescing particles. Data show initial fluorescence rates from time course experiments in epifluorescence mode (from the first frame with 3 sec integration time). **Abbreviations**: GNP – gold nanoparticle; Tf – Transferrin.

Combined populations give a mean $I_B$ of 146 pe s$^{-1}$, which we use in the following. Fluorophores have been shown to have a distribution of bleaching step sizes[36]. Plasmonic NP are known to alter fluorescence intensity depending on the distance of the fluorophore to the NP[37]. Fluorophore quenching has been shown to affect a fluorophore to a distance of up to 15 nm for GNP of 10 nm diameter with a steep decrease in its effects the larger the distance between fluorophore and GNP[38]. The GNPs utilized in our experiments come with a proprietary polymer corona to stabilize them against aggregation in cellular environments for *in vitro* applications, that may prevent or reduce fluorophore quenching due to the GNP. The exact nature of this polymer is not disclosed by the vendor but general *in vitro* studies of GNP make use of PEGylation at molecular weights of 500-5000 kDa, resulting in a polymer thickness of 4-16 nm[39,40]. In addition to the polymer thickness, the transferrin in Tf acts as spacer between the polymer coating and the fluorophore, adding up to 4 nm distance[41] depending on how Tf attaches to the polymer and the location of the fluorophore on transferrin. The resulting total distance of 8-20 nm from the GNP limit the effect of quenching on the detected rates. Additionally, excitation rates are also affected and in general enhanced in the vicinity of GNPs due to enhanced local electric field. Thus, while it is somewhat surprising that the observed bleaching step sizes are similar for free Tf and GNP Tf, it is possible within our mechanistic understanding.

### 3.2 Determination of number of fluorophores per Tf and number of Tfs per GNP

As stated in the previous section, we determine the number of fluorophores $N_F$ by dividing the fluorescence rate (fluorescence intensity spatially integrated over the PSF area of an individual spot in the image) with the photobleaching step size. Figure 2 A shows the histogram of fluorescence rates on a sample containing only free Tf. To calculate $N_F$ we use the average photobleaching step size of 146 pe s$^{-1}$ per fluorophore, and the binning is centred at these step sizes. Transferrin conjugated to a single fluorophore shows the highest occurrence, and it should be noted that transferrin conjugated to zero fluorophores is not detected. The average number of fluorophores per protein $\overline{N}$ was determined by fitting the distribution to a Poissonian statistics yielding $\overline{N}$=1.17 (see Supplementary Figure S3). The manufacturer states an Alexa 647 labelling of 1-3 fluorophores per transferrin, which is consistent with the above value of $\overline{N}$. Panels B-D of Figure 2 show the histogram of $N_F$ from the fluorescence of samples 1 to 3 respectively, with the values originating from Tf no colocalised with a GNP shown in light blue and GNP-Tf constructs in dark blue. Sample 1 and 2 (Figure 2 B and C) show a large distribution of fluorescence rates from GNP-Tf constructs. In both cases, the ratios of attachment site to ligand of 1:0.01 and 1:0.4 during functionalisation do not fully saturate the available reaction sites and therefore increase the probability of two GNPs conjugating to the same Tf. This is supported by the larger fraction of constructs with a fluorescence rate corresponding to more than 15 fluorophores, 15% in sample 1 and 26% in sample 2, compared to 7% in sample 3. The difference between samples 1 and 2 relates to their different sizes of GNPs (14 nm and 20 nm), providing about twice the surface area and thus ligand attachment sites in sample 2. The small fraction of 7% in sample 3 indicates suppressed crosslinking of GNPs. It should be noted that the GNP construct samples used for these experiments were purified by centrifugation, the common method for MNP purification after functionalization with biomolecules[42]. The fraction of fluorescence not colocalised with GNPs is high in all samples (42%, 71%, and 72%) despite purification. This might have serious implications for the evaluation of fluorescence microscopy-based data sets, as signals visible in fluorescence microscopy images would originate from free or aggregated proteins while nominally being attributed to GNP constructs. It is unclear if these free proteins derive from insufficient purification or if they detached over time after the functionalization reaction.

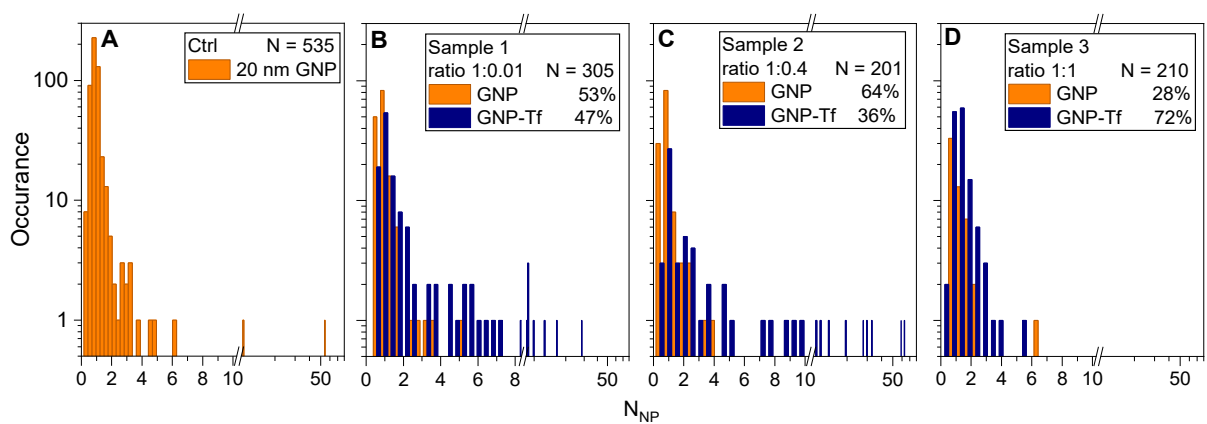

**Figure 3 Extinction cross section of unconjugated GNPs and GNP-Tf constructs.** Histogram of $N_{NP}$ measured on unconjugated 20 nm diameter GNPs prior to functionalisation (A) and on GNP-Tf constructs in sample 1 (B) sample 2 (C), and sample 3 (D). For the GNP constructs, the results have been separated into GNPs without (GNP, orange) and with (GNP-Tf, dark blue) fluorescence, based on the the correlative fluorescence-exciton image analysis. $N_{NP}$ was determined using single GNP extinction cross-sections $\sigma_{sp} = 156$ nm² in (B), and 566 nm² in A, C, and D.

### 3.3 Determination of particle size distribution and crosslinking of particles

We have shown in our previous work[32] that the size of MNPs can be determined quantitatively by measuring the extinction cross section σ$_{ext}$ of individual NPs. We have applied this method here on the same GNP-Tf construct samples imaged by wide-field fluorescence (see Methods section 2.2.2) to verify the presence and size of individual GNPs. Moreover, we determined the number of particles $N_{NP}$ contributing to an extinction signal, in the case of NP multimers (dimers, trimers, etc.) within the PSF, from the ratio between the measured σ$_{ext}$ and the average cross section value of a single particle σ$_{sp}$.

$$N_{NP} = \frac{\sigma_{ext}}{\sigma_{sp}}$$

According to Mie theory[44], GNPs of 14 nm and 20 nm diameter have an extinction cross section of 179 nm² and 547 nm², respectively, at 550 nm excitation wavelength in silicone oil (n = 1.52). The measured σ$_{ext}$ of individual GNPs has a mean value of 156 ± 85 nm² in sample 1, and 566 ± 374 nm² in the other samples, consistent with the nominal diameter from the manufacturer. Figure 3 shows the histograms of $N_{NP}$ measured on the GNP-Tf construct samples (Figure 3 B-D) and on a control sample using unconjugated 20 nm GNPs prior to functionalization (Figure 3 A). The results have been separated into GNPs without (GNP, orange) and with (GNP-Tf, dark blue) fluorescence, from the correlative fluorescence-exciton image analysis described in section 2.3. We find that half to three-quarters of GNPs in all construct samples have no fluorescence (Figure 4 B-D) despite the excess of protein to attachment site and amount of unconjugated protein still present in the sample (Figure 4B-D). A possible reason could be due to the bonds formed by sulfo-SMCC being not stable and shifted toward dissociation with every step of purification. Sulfo-SMCC is a heterobifunctional reagent which is frequently used for crosslinking proteins with an *N*-hydroxylsuccinimide polyethylene (NHS) end on one side of the reagent and a maleimide group on the other. The NHS ester can form stable amide bonds with the primary amine groups on the GNP. In a pH range of 6.5 to 7.5 the maleimide moiety reacts with sulfhydryls in the protein but loses specificity at pH >7.5 which is why the pH should be closely monitored during reaction[43]. Another explanation could be that free Tf forms a protein corona around the constructs that is bound strong enough to stay associated with the construct (soft corona[44,45]) during purification by centrifugation, but not during spin coating when samples are prepared for imaging due to the strong sheer flow. There is a similar variability in the distribution of multimer particles ($N_{NP} > 1$) in sample 1 and 2 (Figure 3 B,C) with significantly less multimers in sample 3 (Figure 4 D), consistent with the previous suggestion of inter-particle crosslinking depending on the ratio of ligands to attachment sites on the particles. In sample 3, the fraction of unfunctionalized particles is lower (28%) compared to sample 1 and 2 (53% and 64%, respectively) suggesting a higher functionalization probability with higher excess of Tf to GNP, as generally

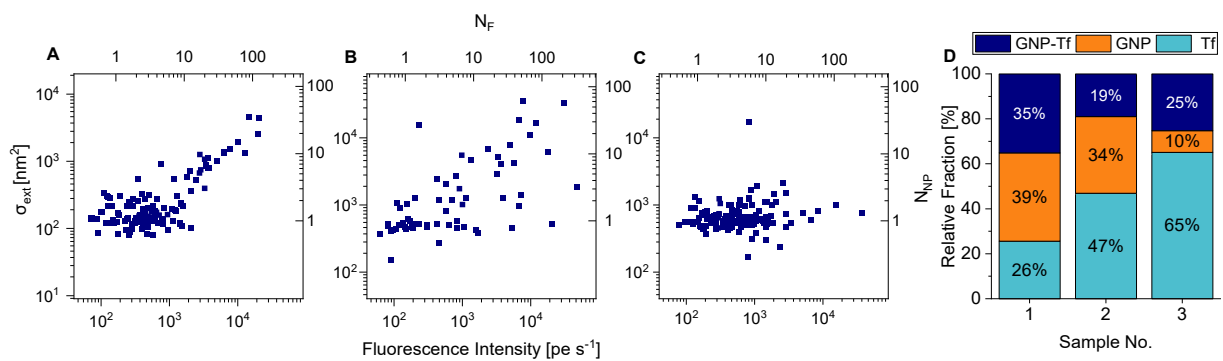

**Figure 4 Characteristics of construct samples.** Correlation of extinction cross-section and fluorescence rates for GNP-Tf constructs from (A) sample 1, (B) sample 2, and (C) sample 3. Larger extinction cross sections indicate a higher number of crosslinked particles while higher fluorescence rates indicate a higher number of bound Tf. D Fractions of unbound protein (light blue), unconjugated particles (orange), and successfully Tf-conjugated GNPs (dark blue) of each sample.

recommended in functionalization protocols of the crosslinkers.

### 3.4 Correlation of fluorescence and extinction signals

More information on the functionalization, beyond the histograms shown can be extracted from the correlative analysis of fluorescence and extinction signals (see Methods section 2.3), by evaluating the extinction cross-sections against fluorescence rates on a single particle basis, as shown in Figure 4 A-C. From this correlation, it can be noted that sample 1 (Figure 4 A), which used the smallest ratio of Tf to GNP, shows a distribution of $\sigma_{ext}$ in the 100-400 nm$^2$ range with fluorescence rates in the 100-1000 pe s$^{-1}$ range. The distribution then narrows into a linear dependence with increasing $\sigma_{ext}$ values. Considering the size distribution of the GNPs (as seen in Figure 3 B), $\sigma_{ext}$ in the 100-400 nm$^2$ range is indicating single GNPs which appear to have a wide spread of fluorescence rates, suggesting a distribution in the range of 1 to 10 fluorophores and thus number or Tf attached per single particle ($\sigma_{ext}$ and fluorescence rates in Figure 4 are additionally shown in units of $\sigma_{sp}$ and bleaching steps, to indicate the number of GNPs and fluorophores). We note that the variation of the fluorescence rate per fluorophore can be significant, specifically when bound to a GNP, so that the actual distribution of fluorophore number per GNP is likely narrower. With increasing $\sigma_{ext}$, indicating GNP multimers in the PSF (likely due to NP cross-linking), the distribution becomes narrower due to averaging of the single particle variabilities across the multimer (ensemble averaging). A different behaviour is observed for sample 3 (Figure 4 C) where $\sigma_{ext}$ is relatively constant indicating mostly single particles, again with a wide variability in fluorescence rates (100-2000 pe s$^{-1}$, indicating 1-15 fluorophores). Hence this sample shows a suppression of GNP crosslinking, but still a high variability in the number of attached fluorophores per nanoparticle. Sample 2 shows a behaviour which is a combination of what observed in the other samples (Figure 4 B), namely a population of single GNPs conjugated to different numbers of Tf, as well as a linear correlation for crosslinked particles. Figure 4 D shows for comparison the overall fractions of unconjugated Tf, conjugated GNP-Tf constructs, and unconjugated GNPs in the samples. From this comparison, we observe that increasing the concentration of Tf during reaction steps reduces the presence of unconjugated GNPs. From an application perspective, this is a useful outcome since unconjugated GNPs are undesirable in targeted cell uptake experiments as they might contribute to non-specific interactions[46]. On the other hand, considering the high abundance of unconjugated protein especially in sample 3, it is surprising that the fraction of conjugated constructs is not higher. A possible reason for this may be hydrolysis of the sulfo-SMCC crosslinking agent which is known to hydrolyse quickly in aqueous solutions. This can be reduced by performing the reaction at low temperatures[47]. Possibly, hydrolysis occurred before there was sufficient time for all particles to find a reaction partner. As already mentioned, another reason for the large fraction of unconjugated protein might be the tendency of proteins to form a soft corona around nanoparticles[44,48] that is then separated from the construct during spin coating. Furthermore, there might be a risk of trapping proteins between NPs during centrifugation. It is nevertheless clear that sample purification which enables a pure, stable population of functionalised nanoparticles must be optimised before cell internalisation studies are performed. The cross-correlation analysis developed here presents an opportunity to assess purification strategy success across conjugated samples before their use.

### 4. Conclusions

We have developed a novel strategy for the quantitative analysis of the functionalization of GNPs decorated with fluorescently labelled proteins on a particle-by-particle basis, by means of optical microscopy on GNP-protein constructs sparsely deposited onto a glass surface. The number of proteins per particle is assessed using widefield fluorescence microscopy combined with an analysis of the fluorescence bleaching step from a single fluorophore, which enables to quantify the number of fluorophores per single protein, and in turn the number of proteins per single particle. The presence, size and number of GNPs is quantitatively characterised using transmission extinction microscopy. The cross-correlative analysis of fluorescence and extinction microscopy gives a quantitative measure of the fraction of single GNPs conjugated to fluorescent proteins, versus unconjugated proteins and GNPs, and the formation of GNP multimers suggesting crosslinked particles. From our results we were able to identify which steps of the conjugation reaction need further improvement. Albeit demonstrated here with spherical GNPs and transferrin proteins, our technique is widely applicable to plasmonic nanostructures of various shapes and materials functionalized with fluorescent biomolecules. Notably, the method is based on relatively simple and easy-to-use widefield microscopy instrumentation, lending itself for widespread adoption of the technique to improve the characterisation of nano-formulation systems.

## Author Contributions

Conceptualization, N. S., P. D. W., P. B., and W. L.; Data curation, N. S.; Formal Analysis, N. S.; Methodology, N. S., P. D. W., P. B., and W. L.; Software, L. P., and W. L.; Validation, W. L., P. B., P. D. W., and N. S.; Investigation, N.S. I.P.; Writing – Original Draft, N. S.; Writing – Review & Editing, L. P., I. P., P. D. W., P. B., W. L.; Visualization, N. S.; Supervision, P. D. W., P. B., W. L.; Funding Acquisition, P. B., Resources P. D. W., and P. B.

## Conflicts of interest

There are no conflicts to declare.

## Acknowledgements

This work was funded by the European Union's Horizon 2020 research and innovation programme under the Marie Sklodowska-Curie grant agreement No 812992. The equipment was funded by the UK Research council EPSRC (grant n. EP/I005072/1 and EP/M028313/1). We acknowledge contributions to the correlation analysis software by Francesco Masia and George Zoriniants.

# Supplementary Information

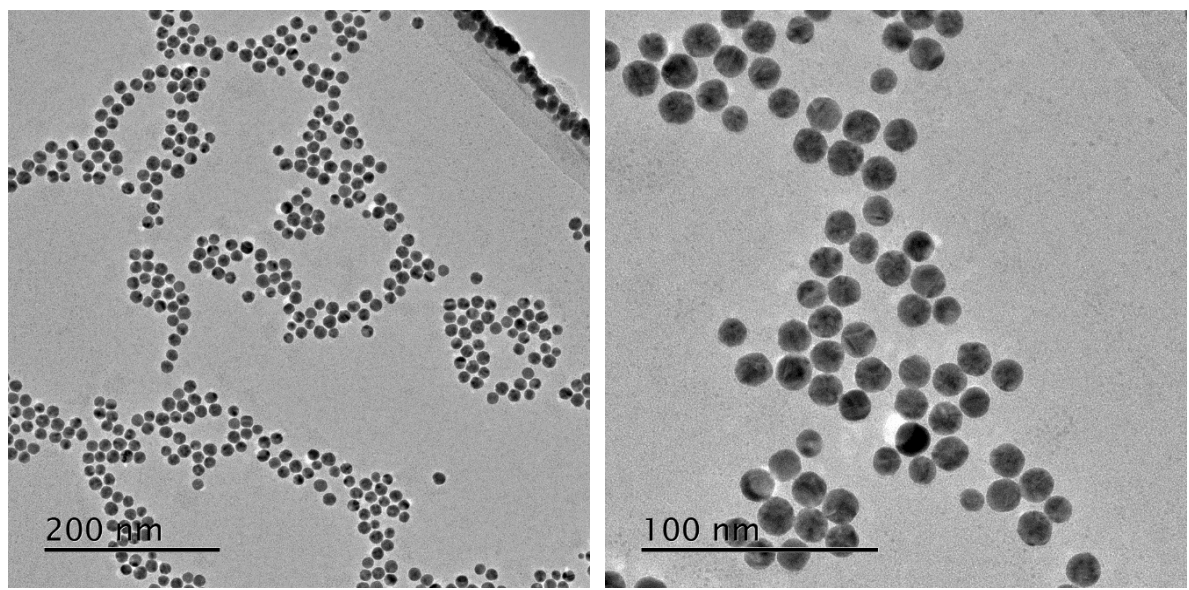

**Figure S1 TEM images of gold nanoparticles (GNPs).** The TEM images show the GNPs used for sample 1 in the main text. 351 particles were analyzed using the particle analysis plugin of imageJ, showing a mean diameter of 13.7 nm and a distribution with a standard deviation of 1.3 nm.

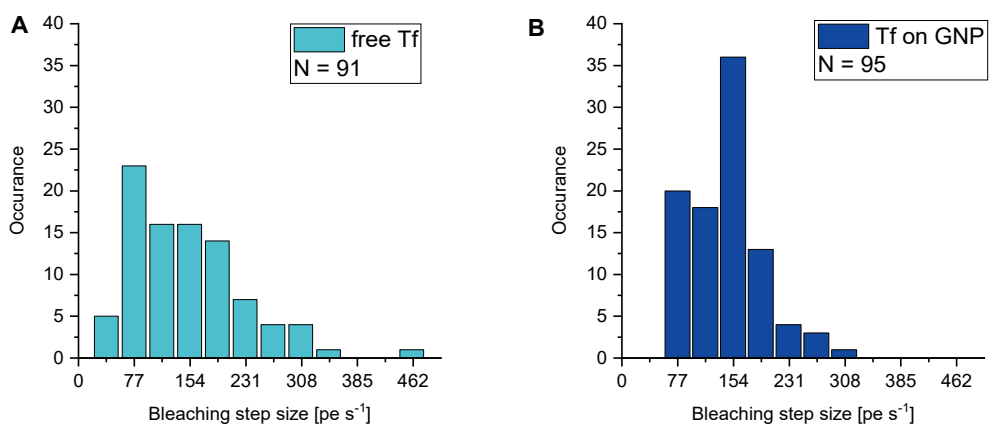

**Figure S2 Bleaching step sizes of Alexa 647.** Histograms of bleaching step sizes $I_B$ of Alexa 647 attached to (A) free transferrin (Tf) and (B) transferrin conjugated to GNPs (GNP-Tf). Tf showed a $I_B$ (mean ± standard deviation) of 150 ± 79 photo-electrons per second (pe s$^{-1}$), and GNP-Tf of 142 ± 48 pe s$^{-1}$.

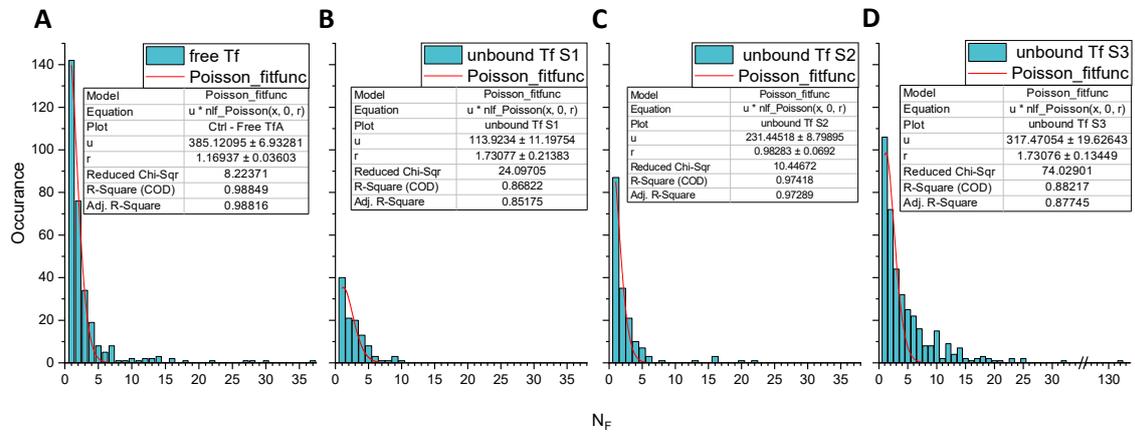

**Figure S3 Histograms of fluorescence rates of Tf.** Distributions of $N_F$ of (A) free Tf, unbound Tf in (B) sample 1, (C) sample 2, and (D) sample 3. The distributions also shown in the main text Figure 2 were fitted by a Poissonian statistics, yielding a mean $N_F$ of 1.17 in the control, 1.73 in sample 1, 0.98 in sample 2, and 1.73 in sample 3. The distribution of sample 3 shows significant deviations from the fit, indicating relevant aggregation of Tf.